\documentclass{article}
\usepackage{spconf,amsmath,amssymb,amsfonts,graphicx}

\usepackage{balance}
\usepackage{algorithmic}
\usepackage[linesnumbered,ruled,vlined]{algorithm2e}

\SetCommentSty{mycommfont}
\usepackage[dvipsnames]{xcolor}
\usepackage{xfrac}
\usepackage{graphicx}
\usepackage{textcomp}
\usepackage{xcolor}
\usepackage{pgfplots, pgfplotstable}
\usepackage{tikz}
\usetikzlibrary{spy} 
\usetikzlibrary{matrix, positioning, patterns, shapes, arrows}
\usetikzlibrary{positioning}
\usepackage{tikzscale}
\usepgfplotslibrary{statistics}
\usepackage{bchart}
\pgfplotsset{compat=newest}
\usepgfplotslibrary{groupplots}
\usepgfplotslibrary{colorbrewer}
\usepackage[linesnumbered,ruled,vlined]{algorithm2e}

\usepackage{amsthm}
\usepackage{thmtools}
\newtheorem{property}{Property}

\declaretheoremstyle[bodyfont=\normalfont]{normalbody}

\graphicspath{{figs/}}

\newcommand\scalemath[2]{\scalebox{#1}{\mbox{\ensuremath{\displaystyle #2}}}}

\title{Recursive/Iterative {unique} Projection-Aggregation decoding of Reed-Muller codes}
\name{Marzieh Hashemipour-Nazari \quad Renate Debets \quad Kees Goossens \quad Alexios Balatsoukas-Stimming}
\address{Eindhoven University of Technology, The Netherlands}
\begin{document}
%
\maketitle
\begin{abstract}
We describe recursive unique projection-aggregation (RUPA) decoding and iterative unique projection-aggregation (IUPA) decoding of Reed-Muller (RM) codes, which remove non-unique projections from the recursive projection-aggregation (RPA) and iterative projection-aggregation (IPA) algorithms respectively. We show that these algorithms have competitive error-correcting performance while requiring up to 95\% projections lower than the baseline RPA algorithm.
\end{abstract}
\section{Introduction}
\label{sec:intro}
While fifth generation networks for mobile telecommunications (5G) are being rolled out, visions for the sixth generation (6G) are already taking shape. The 6G standard will host connections between a wide range of devices with a variety of requirements~\cite{6G}.
For a large number of these mobile devices, it will be essential to have low-latency and high-reliability links. Examples of such applications can be found in autonomous driving, medical monitoring, or other situations where human lives might depend on the speed and robustness of a communication protocol. The protocols used in these cases are ultra reliable low latency communications (URLLC) protocols, and will require the usage of a strong error correction code (ECC) to ensure high reliability. 
However, due to the requirement of low latency for URLLC, transmitted blocks need to be short and finding an ECC scheme that performs well in this short blocklength regime is challenging~\cite{ECC}.

Reed-Muller (RM) codes \cite{Reed1954} have attracted attention recently due their effectiveness for short blocklengths with the development of a near-capacity-achieving decoder called recursive projection-aggregation (RPA)~\cite{Ye2020}. The drawback of this decoder is the high complexity due to its recursive nature, which makes RPA decoding unsuitable for practical implementation. Multiple attempts to reduce its complexity have already been done, for example by collapsing several levels of recursive projection and aggregation into a single level using a technique called collapsed projection-aggregation (CPA)~\cite{Lian2020}, by removing internal iterations resulting in iterative projection-aggregation (IPA) decoding~\cite{IPA}, by reducing the number of branches created during projection~\cite{Fathollahi2021,JiaJie2021,Hashemipour2022}, or by applying a combination of these techniques~\cite{Huang2022}. 

In this work, inspired by~\cite{Lian2020}, we identify non-unique projections in RPA and IPA decoding and we describe decoding algorithms that remove them, namely recursive unique projection-aggregation (RUPA) decoding and iterative unique projection-aggregation (IUPA) decoding. RUPA has the same error-correcting performance as RPA with significantly reduced complexity, while IUPA is competitive with CPA and has potential hardware complexity advantages.
In addition, to further reduce the complexity, the proposed optimization methods for CPA in \cite{Huang2022} and \cite{Li2022} can also be applied to IUPA.


\section{Reed-Muller Codes and Projection-Aggregation-Based Decoding}
\label{sec:BG}

\subsection{Introduction to vector spaces}

Since RM codes form a vector space, we first briefly explain the concepts of vector spaces, subspaces, and cosets that are helpful to study the properties of projection-aggregation-based decoding algorithms~\cite{moon2005}. Consider the binary $m$-dimensional vector space $\mathbb{E}:= \mathbb{F}^{m}_{2}$ with $2^m$ elements, each of which is represented with a binary vector  $\boldsymbol{z} = \left(z_1,\dots,z_m\right)$.
Hence, $\binom{m}{r}$ monomials of degree $r$ can be defined in $\mathbb{E}$. 
Let $\boldsymbol{y}=(y(\boldsymbol{z}),\boldsymbol{z}\in \mathbb{E})$ be a binary incidence vector  of a monomial $F$ defined as:
\begin{equation}\label{eq:monom}
y(\boldsymbol{z}) = f(z_1,\dots,z_m) \forall z\in \mathbb{E},
\end{equation}
where $f(z_1,\dots,z_m)$ is a binary function evaluating monomial $F$ on the element $z\in \mathbb{E}$. In addition, let $\mathbb{B}$ be an $s$-dimensional subspace of $\mathbb{E}$, consisting of $2^s$ elements. The quotient space $\mathbb{E}/ \mathbb{B}$ consists of the following $2^{m-s}$ cosets:
\begin{equation}\label{eq:qoutient}
\mathbb{E}/ \mathbb{B}  = \lbrace T:= \boldsymbol{z}\oplus\mathbb{B}, \boldsymbol{z}\in \mathbb{E}-\mathbb{B} \rbrace,
\end{equation}
where $\oplus$ denotes the binary XOR operation. The projection of the binary vector $\boldsymbol{y}$ onto the cosets of $\mathbb{B}$ is defined as:
\begin{equation}
\label{eq:proj}
\boldsymbol{y}_{/ \mathbb{B}}=\operatorname{Proj}(\boldsymbol{y}, \mathbb{B}):=\left(y_{/ \mathbb{B}}(T):= {\oplus}_{z\in T}y(z), T  \in \mathbb{E} / \mathbb{B} \right).
\end{equation} 
Hence, $\boldsymbol{y}_{/ \mathbb{B}}$ is a $2^{m-s}$-bit binary vector, each coordinate of which is built by summing up the coordinates of $\boldsymbol{y}$ indexed by the elements of each coset $T\in \mathbb{E}/ \mathbb{B}$.
Moreover, ${\binom{m}{s}}_{q=2}$ different $s$-dimensional subspaces are defined for $\mathbb{E}$, where ${\binom{m}{s}}_q$ is the $q$-binomial coefficient and given by:
\begin{equation}
{\binom{m}{s}}_q =\prod_{i=0}^{s-1} \frac{1-q^{m-i}}{1-q^{i+1}}.
\end{equation}
Consequently, ${\binom{m}{s}}_{2}$ different $(2^{m-s})$-bit projected vectors can be defined for a binary $n$-bit vector $\boldsymbol{y}$.

\subsection{Reed-Muller codes}
The $r$-th order Reed-Muller code with blocklength $n = 2^m$ is denoted by RM$(m,r)$.  
The generator matrix $\mathbf{G}_{(m,r)}$ for the RM$(m,r)$ code  consists of incidence vectors of all monomial with a maximum degree $r$. Since $k=\sum _{i=0}^r \binom{m}{i}$ such monomials can be defined in $\mathbb{E}$, the dimension of $\mathbf{G}_{(m,r)}$ is $k\times n$, and the code rate is $R= \frac{k}{n}$.
Similar to other linear block codes, a codeword $\boldsymbol{c}\in \text{RM}(m,r)$ is obtained as $\boldsymbol{c} = \boldsymbol{u}\mathbf{G}_{(m,r)}$, where $\boldsymbol{u}$ is a binary vector with $k$ information bits. Therefore, any codeword $\boldsymbol{c}\in \text{RM}(m,r)$ is the incidence vector of an $m$-variate polynomial with a degree less than or equal to $r$ evaluated for all elements of  $\mathbb{E}$. In addition, it can be shown that the projection of $\boldsymbol{c}$ onto the cosets of the $s$-dimesional subspace $\mathbb{B}$, denoted by $\boldsymbol{c}_{/ \mathbb{B}}$, is a codeword of the RM$(m-s,r-s)$ code~\cite{Ye2020}. 
The RPA and CPA decoding methods utilize this property of RM codes using one-dimensional and $(r-1)$-dimensional subspaces, respectively.
 

\begin{figure}[t]
  \centering
  \centerline{\includegraphics[width=0.49\textwidth]{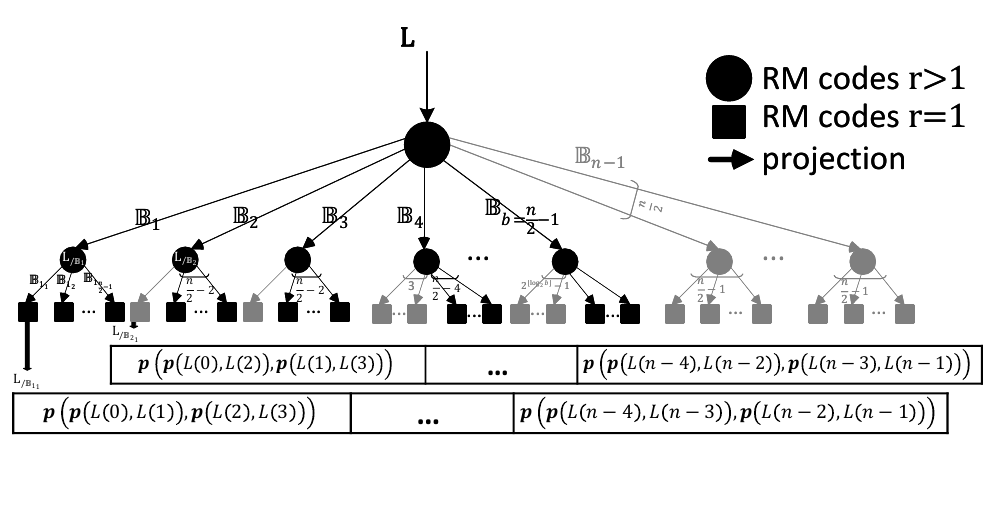}}
  \vspace{-0.75cm}
  \caption{\small RUPA decoding for $r=3$ RM codes.}
  \label{fig:RPA}
\end{figure} 

\subsection{Recursive projection-aggregation (RPA) decoding}

The RPA algorithm decodes a received log-likelihood ratio (LLR) vector $\boldsymbol{L}$ that is obtained from a corrupted version of the the transmitted vector $\boldsymbol{c}\in\text{RM}(m,r)$ in three steps, namely, \emph{projection}, \emph{recursive decoding}, and \emph{aggregation}. 

In the projection step, the received vector is mapped to ${\binom{m}{1}}_{2}=n-1$ new vectors by taking projections on one-dimensional subspaces of an $m$-dimensional binary vector space $\mathbb{E}$. Each one-dimensional subspace $\mathbb{B}_i$ consists of two elements: $\boldsymbol{0}$ and $\boldsymbol{i}$.
The projection rule \ref{eq:proj} for LLR values is:
\begin{equation}
\label{eq:tanh}
\scalemath{0.9}{
\boldsymbol{L}_{/ \mathbb{B}}=\left({L}_{/ \mathbb{B}}(T):=2\tanh^{-1}\left(\prod_{z\in T} \tanh\left(\frac{L(z)}{2}\right) \right)\right),
}
\end{equation}
which is usually approximated by the hardware-friendly min-sum approximation in practice:
\begin{equation}
\label{eq:minsum}
{L}_{/ \mathbb{B}}(T)=\min_{z\in T}\left\lbrace|L(z)|\right\rbrace \prod_{z\in T} \text{sign}\left(L\left(z\right)\right).
\end{equation}

In the recursive decoding step, the projected vectors ${\mathbf{L}}_{/ \mathbb{B}_{i}},i\in\{1,\dots,n-1\},$ are recursively decoded by RPA for $\text{RM}(m{-}1,r{-}1)$ until first-order RM codes are reached after $(r-1)$ levels, which can be decoded using an optimal decoder based on the fast Hadamard transform (FHT)~\cite{Beext86}. 

In the aggregation step, all the decoded codewords obtained from the previous step are aggregated into the LLR vector ${\boldsymbol{\hat{L}}}=\left({\hat{L}}(z),z\in \mathbb{E}\right)$ as follows:
\begin{equation}\label{eq:agg}
{\hat{L}}(z):=\frac{1}{n-1}\sum_{i=1}^{n-1}\left(1-2\hat{y}_{/ \mathbb{B}_i}(z \oplus\mathbb{B}_i)\right)L(z\oplus i),
\end{equation}
where $\boldsymbol{\hat{y}}_{/ \mathbb{B}_i}$ is the hard decision of the decoded codeword from the recursive step. Similar to the projection step, there are $(r-1)$ levels of aggregation until $\boldsymbol{\hat{L}}$ for the input vector $\boldsymbol{L}$ is obtained.
RPA decoding repeats its steps by replacing the input vector $\boldsymbol{L}$ by $\boldsymbol{\hat{L}}$ until either the output converges to the input or $N_{\max}$ iterations are performed.

\subsection{Collapsed projection-aggregation (CPA) decoding}
CPA decoding projects the received LLR vector directly onto $(r-1)$-dimensional subspaces $\mathbb{B}_i,i\in\left\{1,\dots,\binom{m}{r-1}_2\right\},$ to obtain first-order RM codes. 
The projection rule expressed in \eqref{eq:tanh} or \eqref{eq:minsum} still applies for CPA, but each coset $T$ contains $2^{r-1}$ elements in CPA instead of two in RPA. 
After decoding the obtained first-order codes, CPA decoding constructs the vector ${\boldsymbol{\hat{L}}}=\left({\hat{L}}(z),z\in \mathbb{E}\right)$ as follows:
\begin{equation}\label{eq:agg:cpa}
\scalemath{0.875}{
{\hat{L}}(z):=\frac{1}{n_P}\sum_{i=1}^{n_P}-1^{\hat{y}_{/ \mathbb{B}_i}(T)}\left(2 \tanh ^{-1}\left(\prod_{\boldsymbol{z}_i \in T - \boldsymbol{z}} \tanh \left(\frac{L\left(\boldsymbol{z}_i\right)}{2}\right)\right)\right)},
\end{equation}
 where $n_P=\binom{m}{r-1}_2$ is the number of $(r-1)$-dimensional subspaces.
The min-sum approximation is applicable to \eqref{eq:agg:cpa}.
CPA decoding proceeds until either $N_\text{max}$ iterations are reached or the output vector $\boldsymbol{\hat{L}}$ converges to the input vector $\boldsymbol{L}$.

\section{Unique Recursive/Iterative Projection Aggregation Decoding}
\subsection{Duplicate projections}
We first provide the following useful property.
\begin{property}\label{prop:1}
Consider a set of $N$ real values $\mathbf{X}$, and a function $p(\cdot)$ operating on $\mathbf{X}$ as follows:\footnote{This property is also valid for the min-sum approximation of $p(\cdot)$.}
\begin{equation}\label{eq:set_tanh}
p(\mathbf{X}) = 2\tanh^{-1}\left(\prod_{x\in \mathbf{X}} \tanh\left(\frac{x}{2}\right) \right)
\end{equation}
The following property holds:
\begin{equation}\label{eq:gx1x2}
p(\mathbf{X})= 2\tanh^{-1}\left(\prod_{k=1}^{K} \tanh\left(\frac{p \left(\mathbf{X}_k \right)}{2}\right) \right)
\end{equation}
where $\mathbf{X}_k,~k\in \{1,\hdots,K\},$ form a partition of $\mathbf{X}$.
\end{property}
\begin{proof}
Let us replace  $p \left(\mathbf{X}_k \right)$ using \eqref{eq:set_tanh} in \eqref{eq:gx1x2} as follows:
\begin{align}
&2\tanh^{-1}\left(\prod_{k=1}^{K} \tanh\left(\frac{2\tanh^{-1}\left(\prod_{x\in \mathbf{X_k}} \tanh\left(\frac{x}{2}\right) \right)}{2}\right) \right)  \\
&=2\tanh^{-1}\left(\prod_{k=1}^{K} \left(  \prod_{x\in \mathbf{X_k}} \tanh\left(\frac{x}{2}\right) \right) \right)\\
&=2\tanh^{-1}\left(\prod_{x\in \mathbf{X}} \tanh\left(\frac{x}{2}\right) \right) = p(\mathbf{X})
\end{align}
\end{proof}
Now let us take a closer look at the projection step in the RPA decoding with the example shown in Fig.~\ref{fig:RPA}. Consider the received codeword $\boldsymbol{y}\in\text{RM}(m,3)$ represented with the LLR values in the vector $\boldsymbol{L}$. 
The first-order codeword $\boldsymbol{L}_{/\mathbb{B}_{1_{1}}}$ shown in Fig.~\ref{fig:RPA} is created by projecting the first projected vector in level $r=2$ (i.e., $\boldsymbol{L}_{/\mathbb{B}_{1}}$) onto the first $1$-dimensional subspace of $\mathbb{F}^{m-1}_2$.
Similarly, the vector $\boldsymbol{L}_{/\mathbb{B}_{2_{1}}}$ obtained by projecting the $\boldsymbol{L}_{/\mathbb{B}_{2}}$ onto the first $1$-dimensional subspace of $\mathbb{F}^{m-1}_2$.
Considering Property~\ref{prop:1} for $N = 4$ and $K=2$ as well as the projection rule for CPA, $\boldsymbol{L}_{/\mathbb{B}_{1_{1}}}$ and $\boldsymbol{L}_{/\mathbb{B}_{2_{1}}}$ are identical and equal to the projected vector created by projecting $\boldsymbol{L}$ onto $2$-dimensional subspace $\mathbb{B}=\{\boldsymbol{0},\boldsymbol{1},\boldsymbol{2},\boldsymbol{3}\}$ for the CPA decoding. Generally, considering Property~\ref{prop:1} with $N=2^{r-1}$ and $K=2^{r-2}$, the number of duplicate first-order codewords obtained after $r-1$ level of projections for a codeword from RM$(m,r)$ in the RPA algorithm is:
\begin{equation}\label{eq:nd}
N_{\text{D}} = N_{\text{T}}-N_{\text{U}}=\prod_{i=0}^{r-2}{\left(2^{m-i}-1\right)}-\prod_{i=0}^{r-2} \frac{1-2^{m-i}}{1-2^{i+1}},
\end{equation}
where $N_{\text{U}}=\binom{m}{r-1}_2$ is the number of projected first-order codewords obtained after one level of projection onto $(r-1)$-dimensional subspaces in the CPA algorithm~\cite{Lian2020}.
Therefore, the CPA decoder generates the unique projected first-order codewords as it makes projections onto the different $(r-1)$-dimensional subspaces of $\mathbb{E}$. In contrast, after $r-1$ levels of projections in RPA, $N_{\text{T}}$ subspaces with dimension of $r-1$ are built, $N_{\text{D}}$ of which are duplicates. 
To avoid generating duplicate first-order codewords in RPA, we propose recursive unique-projection aggregation (RUPA) decoding.

\begin{algorithm}[t]
\SetAlgoLined
\textbf{Input: } $\mathbf{L}, m,b ,N_{\max},\theta $ \\
\textbf{Output: }Codeword $\hat{\boldsymbol{y}}$\\
\eIf{$r==1$}{
$\boldsymbol{\hat{y}}\leftarrow$ FOD($\mathbf{L},n$)

}{
$n \gets 2^m$ \\
$fp = 2^{\lfloor{\log_{2}b}\rfloor}$ \tcp{first projection}
$lp = 2^{m-r+2}-1$ \tcp{last projection}
\For {$j=0:N_{\max}$}{
\For{$i{=}fp$ : $lp$}{
$\mathbf{L}^i \gets${\ttfamily{Projection}}$\left(\mathbf{L},\mathbb{B}_{i}\right)$ \\ 
$\hat{\boldsymbol{y}}^i\leftarrow$ {\ttfamily{RUPA}}$\left(\mathbf{L}^i,m-1,r-1,i,N_{\max} \right)$  \label{line:rec}\\
$\mathbf{L}_{\text{agg}}^i \gets$ {\small{\ttfamily{Aggregation}}}$\left(\mathbf{L},\hat{\boldsymbol{y}}^i,\mathbb{B}_{i}\right)$ \\
}
 $\mathbf{\hat {L}} \gets \frac{\sum_{fp}^{lp}{\mathbf{L}_{\text{agg}}^i}}{lp-fp+1} $\\
\If{$\vert\mathbf{L}\vert-\vert\mathbf{\hat{L}}\vert <\theta \times \vert\mathbf{L}\vert$}
{break \tcp{early-stopping condition}} 
}
$\mathbf{L}\gets \mathbf{\hat{L}}$\\
$\boldsymbol{\hat{y}} \leftarrow \frac{1-\text{sign}(\mathbf{L})}{2}$  \tcp{hard-decision}
}
\caption{{\ttfamily{RUPA}} decoding for RM$(m,r)$ codes}
\label{alg:RuPA}

\end{algorithm}
\subsection{Unique projections selection}
The projection schedule for RUPA described in Algorithm~\ref{alg:RuPA} selects the $1$-dimensional subspaces at each recursion level of RPA decoding such that there is no duplicate first-order codewords at recursion level $d=r-2$.
Instead of projecting onto all subspaces $\mathbb{B}_i,i\in\{1,\dots,2^{m-d}-1\}$ for each generated vector at the $d$-th level of recursion, $0\leq d \leq r-2$, we select the $\mathbb{B}_i,i\in\{2^{\lfloor{\log_{2}b}\rfloor},\dots\,2^{m-r+2}-1\}$, where $b$ is the branch number indicating the position of the current vector in the tree-like diagram of RPA decoding. 
Due to space constraints, we do not provide a proof that only unique projections are considered by RUPA, but this is indeed the case. 
Therefore, the total number of selected projections for RUPA at $d$-th level is $\binom{m-r+2+d}{d+1}_2$.
As a result, $\prod_{i=0}^{r-2} \frac{1}{2^{i+1}-1}$ of all defined projections at level $d=r-2$ for RPA are kept in RUPA to generate the unique first-order codewords.
The grayed-out branches in Fig.~\ref{fig:RPA} correspond to the duplicate projections in RPA decoding that are skipped in RUPA decoding. Interestingly, Fig.~\ref{fig:RPA} also shows that RUPA keeps more projections at the lower level of recursions and prunes more aggressively in the higher levels, which partially explains the effectiveness of the heuristic pruning method we proposed in~\cite{Hashemipour2022}. 

Apart from the different aggregation rules applied in RUPA and CPA, another difference lies in the internal iterations that RUPA performs at each recursion level, which lead to better performance as we will show. RUPA can be modified to skip internal iterations as in \cite{IPA}, which results in iterative unique-projection aggregation (IUPA) decoding, which has lower complexity and is more hardware-friendly.

\subsection{CPA vs. IUPA}
As mentioned earlier, the projection step in the CPA algorithm leads to the same set of first-order codewords as the IUPA algorithm.
However, their hardware implementation cost may vary significantly. 
For example, in a fully-sequential implementation for the CPA decoder, the projection step requires a crossbar that selects the appropriate cosets for the running projection from all $\binom{m}{r-1}_2$ different projections.
Moreover, in CPA decoding all projections are performed on a vector of size $2^m$ to directly get the first-order codewords. On the other hand, the IUPA decoder obtains $\binom{m}{r-1}_2$ first-order codes in $r-1$ levels of projection, where each level performs projections on vectors of decreasing size. Also, in IUPA the projection step in the  $d$-level, $0\leq d \leq r-2$, requires a crossbar selecting $2^{m-r+2}-1$ different projections for the codewords with size of $2^{m-i}$.  
Moreover, the aggregation step in the CPA decoder requires hardware for the min-sum operations that approximately implement \eqref{eq:agg:cpa}, making the aggregation step in the CPA decoder more complicated than the IUPA decoder.
As such, even though they perform the same projections, it is not clear whether CPA or IUPA will result in the more efficient hardware architecture. It may well be the case that a different algorithm is better suited for different degrees of parallelism, which makes the hardware implementation of IUPA and CPA an interesting and necessary piece of future work.

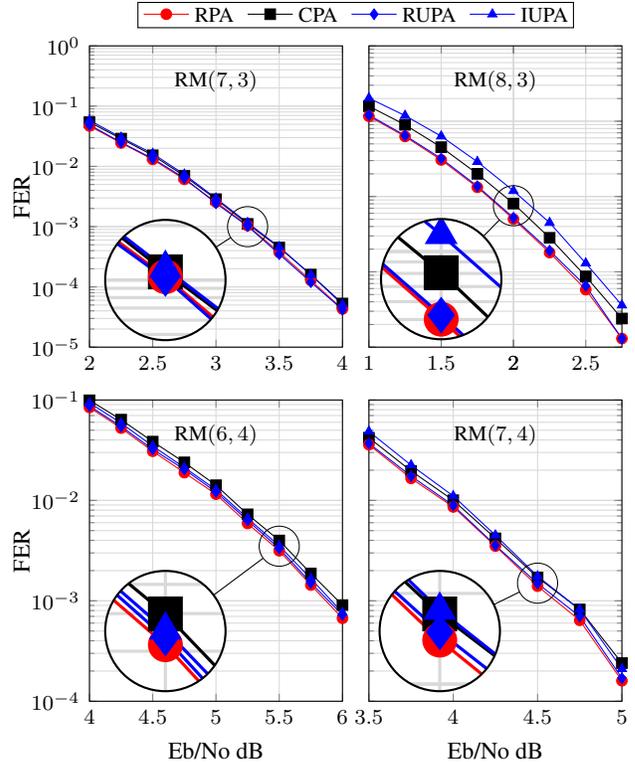
\begin{figure}[t]
	\centering
	\begin{tikzpicture}[spy using outlines={magnification=3,circle,size=1.25cm, black,connect spies}]
		\begin{groupplot}[group style={group name=fer_queries, group size= 2 by 2, horizontal sep=10pt, vertical sep=20pt},
			footnotesize,
			height=.65\columnwidth,  width=.575\columnwidth,
			ymode=log,
			tick align=inside,
			grid=both, grid style={gray!30},
			/pgfplots/table/ignore chars={|},
			]

			\nextgroupplot[ylabel= FER, ytick pos=left, y label style={at={(axis description cs:-0.20,.5)},anchor=south},ymin=1e-5, ymax =1,xmin=2, xmax=4, xtick={2,2.5,3,3.5,4}]

\coordinate (p211) at (axis cs: 3.25,1e-3);
\coordinate (p212) at (axis cs: 2.6,1.3e-4);
\spy[size=1.6cm] on (p211) in node[fill=white] at (p212);

\addplot[ color=black ,mark=square* ] coordinates {
( 2.00, 0.05415358)
( 2.25, 0.02888838)
( 2.50, 0.01527370)
( 2.75, 0.00694000)
( 3.00, 0.00286000)
( 3.25, 0.00111000)
( 3.50, 0.00045000)
( 3.75, 0.000160)
( 4.00, 0.00005300)
};\label{gp:CPA}

\addplot[ color=red ,mark=* ] coordinates {
( 2.00, 0.04686036)
( 2.25, 0.02448820)
( 2.50, 0.01303186)
( 2.75, 0.00609827)
( 3.00, 0.00255429)
( 3.25, 0.00104325)
( 3.50, 0.00037500)
( 3.75, 0.00012800)
( 4.00, 0.00004300)
};\label{gp:RPA}

\addplot[ color=blue ,mark=triangle* ] coordinates {
( 2.00, 0.05828865)
( 2.25, 0.03017502)
( 2.50, 0.01629036)
( 2.75, 0.00736000)
( 3.00, 0.00295000)
( 3.25, 0.00116000)
( 3.50, 0.00046000)
( 3.75, 0.00015900)
( 4.00, 0.00005100)
};\label{gp:UIPA}

\addplot[ color=blue ,mark=diamond* ] coordinates {
( 2.00, 0.04885198)
( 2.25, 0.02500875)
( 2.50, 0.01350202)
( 2.75, 0.00636000)
( 3.00, 0.00244000)
( 3.25, 0.00102000)
( 3.50, 0.00035000)
( 3.75, 0.00012000)
( 4.00, 0.00004200)
};\label{gp:URPA}
\coordinate (top) at (rel axis cs:0,1);


\nextgroupplot[yticklabels=\empty,ymin=1e-4, ymax = 1,xmin=1, xmax=2.75, xtick={1,1.5,2,2,2.5}]

\coordinate (p311) at (axis cs: 2,7.55e-3);
\coordinate (p312) at (axis cs: 1.5,8e-4);
\spy[size=1.6cm] on (p311) in node[fill=white] at (p312);
		
\addplot[ color=black ,mark=square* ] coordinates {
( 1.00, 0.15822034)
( 1.25, 0.08953000)
( 1.50, 0.04518000)
( 1.75, 0.02002000)
( 2.00, 0.00806000)
( 2.25, 0.00283000)
( 2.50, 0.00087000)
( 2.75, 0.00024000)
};

\addplot[ color=red ,mark=* ] coordinates {
( 1.00, 0.11628177)
( 1.25, 0.06267000)
( 1.50, 0.03058000)
( 1.75, 0.01324000)
( 2.00, 0.00502000)
( 2.25, 0.00180000)
( 2.50, 0.00058000)
( 2.75, 0.00013000)
};

\addplot[ color=blue ,mark=triangle* ] coordinates {
( 1.00, 0.20250294)
( 1.25, 0.11860286)
( 1.50, 0.06317000)
( 1.75, 0.02892000)
( 2.00, 0.01189000)
( 2.25, 0.00449000)
( 2.50, 0.00130000)
( 2.75, 0.00036000)
};

\addplot[ color=blue ,mark=diamond* ] coordinates {
( 1.00, 0.12054002)
( 1.25, 0.06460000)
( 1.50, 0.03197000)
( 1.75, 0.01374000)
( 2.00, 0.00523000)
( 2.25, 0.00191000)
( 2.50, 0.00065000)
( 2.75, 0.00013000)
};


			\nextgroupplot[ylabel= FER, ytick pos=left, y label style={at={(axis description cs:-0.20,.5)},anchor=south},ymin=1e-4, ymax = 1e-1,xmin=4, xmax=6, xtick={4,4.5,5,5.5,6},xlabel=Eb\slash No dB]

\coordinate (bot) at (rel axis cs:1,0);

\coordinate (p121) at (axis cs: 5.5,3.5e-3);
\coordinate (p122) at (axis cs: 4.6,5e-4);
\spy[size=1.6cm] on (p121) in node[fill=white] at (p122);

\addplot[ color=black ,mark=square* ] coordinates {
( 4.00, 0.09932459)
( 4.25, 0.06366183)
( 4.50, 0.03848818)
( 4.75, 0.02390514)
( 5.00, 0.01415168)
( 5.25, 0.00726639)
( 5.50, 0.00399017)
( 5.75, 0.00187422)
( 6.00, 0.00090700)
};

\addplot[ color=red ,mark=* ] coordinates {
( 4.00, 0.08419635)
( 4.25, 0.05243838)
( 4.50, 0.03086229)
( 4.75, 0.01890431)
( 5.00, 0.01153256)
( 5.25, 0.00589018)
( 5.50, 0.00315145)
( 5.75, 0.00143552)
( 6.00, 0.00067000)
};

\addplot[ color=blue ,mark=triangle* ] coordinates {
( 4.00, 0.09363296)
( 4.25, 0.05880278)
( 4.50, 0.03478140)
( 4.75, 0.02127433)
( 5.00, 0.01278217)
( 5.25, 0.00663420)
( 5.50, 0.00352879)
( 5.75, 0.00166263)
( 6.00, 0.00076400)
};

\addplot[ color=blue ,mark=diamond* ] coordinates {
( 4.00, 0.08697921)
( 4.25, 0.05448404)
( 4.50, 0.03227681)
( 4.75, 0.02036826)
( 5.00, 0.01212209)
( 5.25, 0.00631540)
( 5.50, 0.00332259)
( 5.75, 0.00151122)
( 6.00, 0.00072500)
};

\nextgroupplot[yticklabels=\empty,ymin=1e-4, ymax = 1e-1,xmin=3.5, xmax=5, xtick={3.5,4,4.5,5},xlabel=Eb\slash No dB]

\coordinate (p221) at (axis cs: 4.5,1.5e-3);
\coordinate (p222) at (axis cs: 3.92,5e-4);
\spy[size=1.6cm] on (p221) in node[fill=white] at (p222);

\coordinate (bot) at (rel axis cs:1,0);
\addplot[ color=black ,mark=square* ] coordinates {
( 3.50, 0.04206099)
( 3.75, 0.01967536)
( 4.00, 0.01003291)
( 4.25, 0.00419000)
( 4.50, 0.00171000)
( 4.75, 0.00082000)
( 5.00, 0.00024000)
};

\addplot[ color=red ,mark=* ] coordinates {
( 3.50, 0.03586157)
( 3.75, 0.01656617)
( 4.00, 0.00861000)
( 4.25, 0.00351000)
( 4.50, 0.00140000)
( 4.75, 0.00064000)
( 5.00, 0.00016000)
};

\addplot[ color=blue ,mark=triangle* ] coordinates {
( 3.50, 0.04863577)
( 3.75, 0.02244619)
( 4.00, 0.01107432)
( 4.25, 0.00450000)
( 4.50, 0.00176000)
( 4.75, 0.00082000)
( 5.00, 0.00021000)
};

\addplot[ color=blue ,mark=diamond* ] coordinates {
( 3.50, 0.03700688)
( 3.75, 0.01747213)
( 4.00, 0.00891000)
( 4.25, 0.00357000)
( 4.50, 0.00149000)
( 4.75, 0.00070000)
( 5.00, 0.00017000)
};


		\end{groupplot}
		\node[below = 0.2cm of fer_queries c1r1.north] { \footnotesize $\text{RM}(7,3)$ };
		\node[below = 0.2cm of fer_queries c2r1.north] {\footnotesize $\text{RM}(8,3)$};
		\node[below = 0.2cm of fer_queries c1r2.north] { \footnotesize $\text{RM}(6,4)$ };
		\node[below = 0.2cm of fer_queries c2r2.north] {\footnotesize $\text{RM}(7,4)$};
					
		\path (top|-current bounding box.north) -- coordinate(legendpos) (bot|-current bounding box.north);
		\matrix[
		matrix of nodes,
		anchor=south,
		draw,
		inner sep=0.1em,
		draw,
		column 1/.style={anchor=base west},
    	column 2/.style={anchor=base west},
    	column 3/.style={anchor=base west},
    	column 4/.style={anchor=base west},
		]at(legendpos)
		{
			\ref{gp:RPA}& \footnotesize RPA  &[3pt]
			\ref{gp:CPA}& \footnotesize CPA  &[3pt]
			\ref{gp:URPA}& \footnotesize RUPA &[3pt]
			\ref{gp:UIPA}& \footnotesize IUPA \\
			};
	\end{tikzpicture}  
	
	\caption{FER comparison between RPA, CPA, RUPA, and IUPA decodings for different RM codes.}
	\label{fig:result}
\end{figure}

\section{Results}
\label{sec:res}
In this section, we compare the error-correcting performance of the proposed RUPA and IUPA algorithms to RPA decoding and CPA decoding.
We present simulation results over an additive white Gaussian noise (AWGN) channel for RM$(7,3)$, RM$(8,3)$, RM$(6,4)$ and RM$(7,4)$ codes to explore the effect of the proposed projection selection on the error-correcting performance of RM codes of various blocklengths and rates.
We set $N_{\max}=3$ for codes with $m=6$ and $m=7$ and $N_{\max}=4$ for RM$(8,3)$ code.
Moreover, we use the hardware-friendly min-sum approximation for the projection rule in \eqref{eq:proj} and aggregation rule in \eqref{eq:agg:cpa}. 
 
As shown in Fig.~\ref{fig:result}, there is effectively no performance difference between RUPA decoding and RPA decoding, even though RUPA only keeps $\sfrac{1}{3}$ and $\sfrac{1}{21}$ of the all projections performed in RPA for third-order and fourth-order codes, respectively. As a result, RUPA decoding can close the gap to RPA decoding that CPA decoding exhibits for RM$(8,3)$.

In addition, the difference in performance between CPA and IUPA is negligible for RM$(6,4)$, RM$(7,3)$, and RM$(7,4)$ codes. However, a performance loss of $0.10$~dB is observed for IUPA compared to CPA for RM$(8,3)$. Therefore, the aggregation rule employed in the CPA algorithm for the larger codes has a more positive effect on its overall decoding performance compared to the IUPA algorithm. However, as described previously, using the min-sum operation in the aggregation step of the CPA decoder may translate to a more complicated hardware implementation and it is still unclear whether the trade-off is worth it.

\section{Conclusion}
In this work, we proposed a schedule for selecting the unique projections in the RPA decoding method, resulting in the RUPA decoding algorithm. Our simulation results showed that RUPA decoding for RM$(m,r)$ does not degrade the performance of the baseline RPA decoding while reducing the required computations by up to 95\% for the examined RM codes.
In addition, we removed the internal iterations in RUPA based on the observation in \cite{IPA}, resulting in the IUPA decoding algorithm that is more suitable for hardware implementation. Moreover, we argued that, although the same first-order codewords are obtained after the projection step in the CPA and IUPA decoders, the hardware implementation efficiency of the two algorithms depends on the degree of parallelization and other implementation details that need to be explored further. Finally, our results show that CPA only performs better than IUPA for certain codes even though it has a more advanced and costly aggregation rule.

\vfill\pagebreak
\balance

\bibliographystyle{IEEEbib}
\bibliography{refs}

\end{document}